# Unusually Strong Four-Phonon Scattering Effects on Low-Temperature Thermal Conductivity in Two-Dimensional Materials

H. F. Feng,[1] B. Liu,[1] Xin-Gao Gong,[2,†] and Zhi-Xin Guo[1,*]

[1]*State Key Laboratory for Mechanical Behavior of Materials, School of Materials Science and Engineering, Xi'an Jiaotong University, Xi'an, Shanxi, 710049, China.*
[2]*Key Laboratory for Computational Physical Sciences (MOE), Institute of Computational Physics, Fudan University, Shanghai 200433, China.*

†xggong@fudan.edu.cn
*zxguo08@xjtu.edu.cn

## Abstract

First principles-based predictions of lattice thermal conductivity (TC) from perturbation theory have achieved significant success. Usually, it only included three-phonon (3ph) scattering processes, only recently four-phonon (4ph) scattering processes were found to have a comparable impact as 3ph scattering at medium and high temperatures in various materials. While the influence of 4ph scattering on TC at low temperatures was generally believed to be insignificant. By combining the first-principles calculations, machine learning techniques, and Boltzmann transport equation (BTE), we find that there are unusually strong 4ph processes even in the low-frequency range of two-dimensional (2D) materials such as h-XN (X = B, Al, Ga), which have a remarkable influence on the low-temperature TC. Such strong 4ph processes originated from the out-of-plane acoustic (ZA) phonon mode of 2D materials. Furthermore, we find that the intensity of 4ph scattering and thus TC can be effectively manipulated by changing the dispersion of ZA phonon mode, which can be easily achieved through strain engineering. The present study provides new insights into low-temperature phonon transport and its manipulation in 2D materials.

Thermal conductivity (TC) plays a crucial role in the performance of materials across a wide range of applications, accurate prediction of lattice thermal conductivity is therefore of great interest to both researchers and engineers [1-9]. First-principles-based approaches, particularly those employing perturbation theory, have made substantial progress in predicting TC by considering phonon scattering processes, with the three-phonon (3ph) scattering mechanism being the primary focus [10-14]. However, recent studies have shown that relying solely on 3ph processes is insufficient for accurate TC estimation and four-phonon (4ph) scattering must be considered for improved accuracy [15-20]. The inclusion of 4ph scattering leads to a significant decrease of TC at medium and high temperatures, which gives rise to remarkable agreement between the theoretical and experimental results [17,21-24].

However, the influence of 4ph scattering at low temperatures has generally been overlooked, since it is believed the contribution of 4ph should be small, owing to the fact that the scattering rates of 3ph and 4ph are proportional to $T\omega^2$ and $T^2\omega^4$, respectively (T represents temperature, $\omega$ indicates frequency. Note that only small ω can be excited at low temperatures) [11,12]. Nevertheless, the out-of-plane acoustic (ZA) phonon mode, uniquely existing in two-dimensional (2D) materials, usually has a remarkable contribution to TC [25,26]. It is noted that ZA phonon mode normally has very low frequencies, which may have a significant influence on TC even at low temperatures from the viewpoint of the Bose-Einstein distribution of phonons. This means that the 4ph processes correlated to ZA mode have to be seriously considered even at low temperatures.

In this work, we combine first-principles calculations[27-29], machine learning potential (MLP) methods[30-32], and the Boltzmann transport equation (BTE) to investigate the role of 4ph scattering in 2D materials [10-12,33], specifically h-XN (X = B, Al, Ga) monolayers. Our findings reveal unusually strong 4ph processes in the low-temperature regime of these materials, with the ZA phonon mode being a major contributor. This leads to a substantial reduction in TC at low temperatures. Furthermore, we demonstrate that the intensity of 4ph scattering, and hence TC, can be effectively manipulated through strain engineering, offering a novel avenue for controlling the thermal properties of 2D materials.

In 2D materials, a unique acoustic phonon mode (ZA mode) exhibits a quadratic dispersion relation near Γ point. This behavior contrasts with that of bulk materials, where all three acoustic phonon modes display linear dispersion near Γ point [34-36]. For example, as shown in Fig. 1(a), graphene (Gr) and bulk Silicon (Si) exhibit distinct acoustic phonons. In Gr, the three acoustic phonon modes are ZA, longitudinal (LA), and transverse (TA) modes, while in bulk Si, only LA and TA modes are present. The TA and LA in both Gr and Si present linear dispersion around Γ point, and the ZA mode in Gr follows an ideal quadratic curve. This quadratic dispersion results in a significantly higher number of excited phonons (NEP) at low frequencies, owing to a larger phonon density of states and Bose-Einstein distribution. Figure 1(b) illustrates the phonon frequency-dependent NEP for Gr and Si. It is evident that in the low-frequency region (<3 THz), the ZA mode dominates the NEP in Gr, with values 10 to 100 times larger than those of LA and TA modes. This dominance indicates the ZA mode makes the primary contribution to the TC of graphene at low temperatures (<100 K). In contrast, in Si, the NEP arises purely from the LA and TA modes, leading to significantly lower NEP values compared to the ZA mode in Gr.

The exceptionally large NEP of the ZA mode in 2D materials contributes to unexpectedly strong 4ph scatterings at low temperatures. Monolayer 2D materials possess reflection symmetry perpendicular to the plane, which imposes a symmetry-based selection rule that prohibits scattering channels with an odd number of ZA modes [26]. Consequently, the three-ZA processes (ZA ↔ ZA+ZA) are forbidden, which lead to weak 3ph scatterings and make the ZA mode contribute greatly to TC, as previously reported [37-40]. Nonetheless, four-phonon processes involving an even number of ZA modes (such as ZA+ZA ↔ ZA+ZA and ZA ↔ ZA+ZA+ZA) remain permitted and lead to strong 4ph scatterings. Considering that at low temperatures, the predominant excitation is the low-frequency phonons and the ZA mode has a dominating NEP in the low-frequency region, the TC in 2D materials is expected to be significantly influenced by the strong 4ph scatterings.

To verify the above arguments, we investigate a family of 2D materials, hexagonal nitrogen-based III–V monolayers (h-XN, X = B, Al, Ga), which exhibit similar geometry but significantly distinct phonon dispersions due to the different atomic mass X. Therefore, a systemic investigation of the influence of 4ph processes on TC in the low-temperature region can be achieved via the study on the h-XN system. We begin by performing first-principles calculations based on the density functional theory (DFT) to obtain the atomic structures of h-XN monolayers. As shown in Figs. 2(a)-2(c), all three h-XN monolayers possess a 2D planar structure. Figures 2(d)-2(f) further show the phonon dispersions (SI in Supplemental Material). A common characteristic among these 2D materials is the existence of a ZA mode with an ideal quadratic dispersion in the low-frequency region, which is widely recognized to make a significant contribution to TC [26].

TC calculations are performed using the BTE method (SI in Supplemental Material for calculation details [41]). The BTE method, derived from the Boltzmann kinetic theory, is a representation of the general Green-Kubo expression for TC calculations [53]. Focusing on the TC of 2D simple crystals allows us to ignore inter-band scattering effects, simplifying the analysis of transport properties (SVI in Supplemental Material). This approach allows us to clearly differentiate between the effects of 3ph processes and 4ph processes. Consequently, the BTE method alone is sufficient to describe our system.

In order to improve the efficiency of calculation in the BTE method, we further construct the MLP using the neuroevolution potential model [30,42], based on the first-principles calculations. Figure S1 shows good consistency of forces and energies between first-principles and MLP calculations. Additionally, we calculate the phonon dispersions of h-XN monolayers based on MLP. As shown in Fig. 2, phonon dispersion predicted by MLP (black solid line) shows excellent agreement with that predicted by the first-principles calculations (red circles) for h-XN systems. These results confirm that the MLP can accurately describe the harmonic properties of the h-XN monolayers, which is a vital prerequisite to describing the energy and momentum conservation in phonon-phonon scattering processes, enabling an accurate prediction of TC.

Next, we investigate the temperature-dependent TC of the h-XN monolayers. Based on the trained MLP, we calculate TC at varying temperatures using the BTE method, considering 3ph ($TC_{3ph}$) and both 3ph and 4ph ($TC_{3+4ph}$) scatterings. As shown in Figs. 3(a)-3(c), regardless of considering only 3ph scattering or both 3ph and 4ph (denoted as 3+4ph) scatterings, TC of the three h-XN monolayers decreases monotonically with temperature increasing from 100 K to 900 K. Note that such monotonic reduction of TC with temperature is consistent with previous studies in the BTE calculations [17,33]. Another common feature is that a notable reduction of TC can be obtained when the 4ph scattering effect is included. In Fig. 3(d) we additionally show the temperature-dependent ratio of $TC_{3+4ph}$ and $TC_{3ph}$ (denoted as $R_{TC}$), which can effectively illustrate the significance of 4ph scattering on TC. Remarkably, it is found that $R_{TC}$ remains quite small ($R_{TC} \leqslant 0.66$) across the whole simulated temperature range for the h-XN monolayers. Particularly, at 100 K, the reduction of TC reaches up to 34%, 79%, and 54% for BN (from 2612 to 1736 W/mK), AlN (from 454 to 95 W/mK), and GaN (from 127 to 58 W/mK), respectively. Especially, the reduction of TC in AlN at 100 K is even more significant than that at 900 K, which is contrary to the general concept, i.e., a more significant reduction always appears at higher temperatures, because 4ph processes are expected to be more significant at high temperatures. Note that, we also consider the influence of rotational invariance and find it has little effect on the results [Fig. S2]. The above finding clearly shows the unusually significant 4ph scattering effect on TC of h-XN monolayers.

It is noticed that $R_{TC}$ in the XN monolayers exhibits a strong dependence on the X element. That is, in BN, it remains almost constant at about 0.63, whereas in AlN and GaN, it initially increases (from 0.21 to 0.36 for AlN, from 0.46 to 0.47 for GaN) and then decreases with increasing temperature from 100 K to 900 K. The stronger 4ph effect on TC in AlN and GaN can be attributed to the larger NEP of out-of-plane phonons in the low-frequency region (see Fig. S3), which usually corresponds to the stronger 4ph processes. Note that both two types of variations of $R_{TC}$ are very different from those in 3D materials, where $R_{TC}$ monotonically decreases with increasing temperature due to the more pronounced enhancement of 4ph processes [58,59]. To confirm our BTE calculations, we have also calculated the temperature-dependent TC [Fig. S4(a)] and $R_{TC}$ [Fig. 3(d)] of bulk Si. Encouragingly, our results demonstrate a noticeable decrease in the 4ph effect as the temperature decreases, aligning well with previous investigations [17]. This phenomenon agrees well with the 3D phonon dispersions and NEP distributions shown in Fig. 1. That is, the absence of ZA mode and thus the significantly smaller NEP in the low-frequency region, in comparison to the 2D structures, give rise to weaker 4ph scatterings with the decrease of temperature.

To understand the underlying mechanisms responsible for the reduction in TC when considering 4ph scattering processes, we analyze the contribution of individual phonon modes to the overall thermal conductivity and its reduction. We focus on h-XN monolayers at 100 K as illustrative examples for a detailed discussion. At this temperature, low-frequency phonons are predominantly excited, making the ZA mode particularly significant. As depicted in Fig. 4(a), calculations considering only 3ph scattering reveal that the ZA mode contributes 2385.8 W/mK, 425.4 W/mK, and 102.7 W/mK to the total TC of h-BN, h-AlN, and h-GaN, respectively. Notably, the ZA mode accounts for over 80% of the total thermal conductivity in these materials when only 3ph processes are considered. However, upon incorporating 4ph scattering, a substantial decrease in total TC is observed, and this reduction is primarily attributed to the ZA mode. To quantify this effect specifically for the ZA mode, we define the ratio ($R_{TC\text{-}ZA}$) of the thermal conductivity reduction due to the ZA mode and $TC_{3ph}$. This ratio is expressed as $\Delta TC_{ZA}/TC_{3ph}$, where $\Delta TC_{ZA} = TC_{3ph-ZA} - TC_{3+4ph-ZA}$. Here, $TC_{3ph-ZA}$ and $TC_{3+4ph-ZA}$ represent the TC contribution from the ZA mode when considering only 3ph and 3+4ph scattering processes, respectively. As depicted in Fig. 4(b), the ZA mode exhibits a significant reduction in its contribution to TC due to 4ph scattering, amounting to approximately 30%, 78%, and 56% for h-BN, h-AlN, and h-GaN, respectively.

Comparing these ZA-mode-induced reductions with the overall reduction in $R_{TC}$ presented in Fig. 3(d), it becomes evident that the pronounced $R_{TC}$ values observed are predominantly attributed to the ZA mode contribution within h-XN monolayers. In contrast, for 3D materials like bulk silicon, LA and TA modes are the primary contributors to TC at low temperatures, as shown in Fig. S4(b). Both LA and TA modes exhibit a considerably smaller reduction in TC when 4ph scattering is considered (Fig. S4(c)). Consequently, 3D materials typically do not display the same pronounced 4ph scattering effects observed in two-dimensional materials at low temperatures.

The above results provide a clear explanation for the anomalously large 4ph effect on TC of 2D materials in the low-temperature region, which comes from ZA mode. Especially, in 2D compounds with distinct atomic masses, such as AlN and GaN, the strong 4ph processes can result in a variation of TC over 67% at 100 K. To find out the underlying reasons for the reduced TC of ZA mode when considering 4ph processes, we calculate the phonon scattering rates for both 3ph and 3+4ph processes, which are inversely proportional to TC [10]. Note that here we only focus on the scattering rates in the low-frequency region ($\omega$ < 3.0 THz), where the phonons are expected to be excited at 100 K. As shown in Figs. 5(a)-5(c), the 3+4ph scattering rates are higher than those of 3ph in the low-frequency region, especially for h-AlN and h-GaN, where these materials exhibit much higher NEP and more pronounced scattering between ZA modes [Fig. S3]. This result clearly shows that the variation of scattering rates has a significant contribution to TC in the low-temperature region.

As indicated in Fig. 2, the low-frequency phonons ($\omega$ < 3.0 THz) are dominated by the three acoustic modes, TA, LA, and ZA modes. In order to further distinguish the 4ph effect of each mode, as shown in Figs. 5(d)-5(f), we plot the ratio of scattering rates (denoted as $R_s$) between 3+4ph processes and 3ph processes for these three acoustic modes. It can be seen that in all the h-XN monolayers, $R_s$ of the ZA mode is larger than 1.0 and surpasses that of LA and TA modes (especially for AlN and GaN). This feature indicates that the unusually large 4ph scattering processes in the low-frequency region mainly come from the ZA mode, which is a distinctive characteristic exclusive to 2D materials. It is worth noting that the ZA mode in AlN/BN has the largest/smallest $R_s$ value, showing the most significant/insignificant 4ph processes in the AlN/BN monolayer. This result is consistent with the variation of $R_{TC}$ with element X in the h-XN monolayers discussed above. On the other hand, we have also calculated $R_s$ of the three acoustic modes (LA, TA) in the bulk Si and found all of them have $R_s$ around 1.0 in the low-frequency region, which shows the insignificant 4ph processes in 3D materials [Fig. S4(e)].

The strong correlation between the 4ph processes and the ZA phonon mode offers a promising avenue for manipulating TC in 2D materials, for example, through the application of tensile strain. Here we adopt the h-AlN monolayer as an illustration. Fig. S5(a) presents that the ZA mode exhibits an obvious upshift and linearization under the tensile strain, exhibiting greater sensitivity compared to LA and TA modes. As illustrated in Fig. S5(b), the linearization of the ZA mode leads to a significant decrease in the NEP of out-of-plane mode in the low-frequency region, which substantially affects the scattering processes involving the ZA mode. This results in a notable change in the scattering rates across the entire frequency range, as shown in Fig. S5(c), which presents the scattering rates for 3ph and 3+4ph processes before and after stretching at 100 K. The variation in the scattering rate for the 3+4ph processes is more pronounced compared to that of the 3ph processes. The temperature-dependent TC with and without strain is further shown in Fig. S5(d). A noteworthy increase of TC considering both 3ph and 3+4ph scatterings has been obtained when a 1% tensile strain is applied. Particularly, at 100 K the increment of TC considering 3+4ph scattering (315%) is much more significant than that considering only 3ph scattering (223%). This feature leads to an enhancement of $R_{TC}$ in the low-temperature region as shown in Fig. S5(e). The above results provide clear evidence that the TC of 2D materials in the low-temperature region can be effectively modulated due to the strong correlation between 4ph processes and the ZA mode.

Before closing, we remark that the strong 4ph processes at low temperatures should widely exist in 2D materials, in which the ZA mode has a significant contribution to TC. This finding provides an avenue for studies on high-order phonon interactions of 2D materials in low-temperature regions.

We are grateful for useful discussions with Professor Wu Li and Zhibin Gao. We acknowledge financial support from the Natural Science Foundation of China (Grant

No. 12474237) and Science Fund for Distinguished Young Scholars of Shaanxi Province (Grant No. 2024JC-JCQN-09).

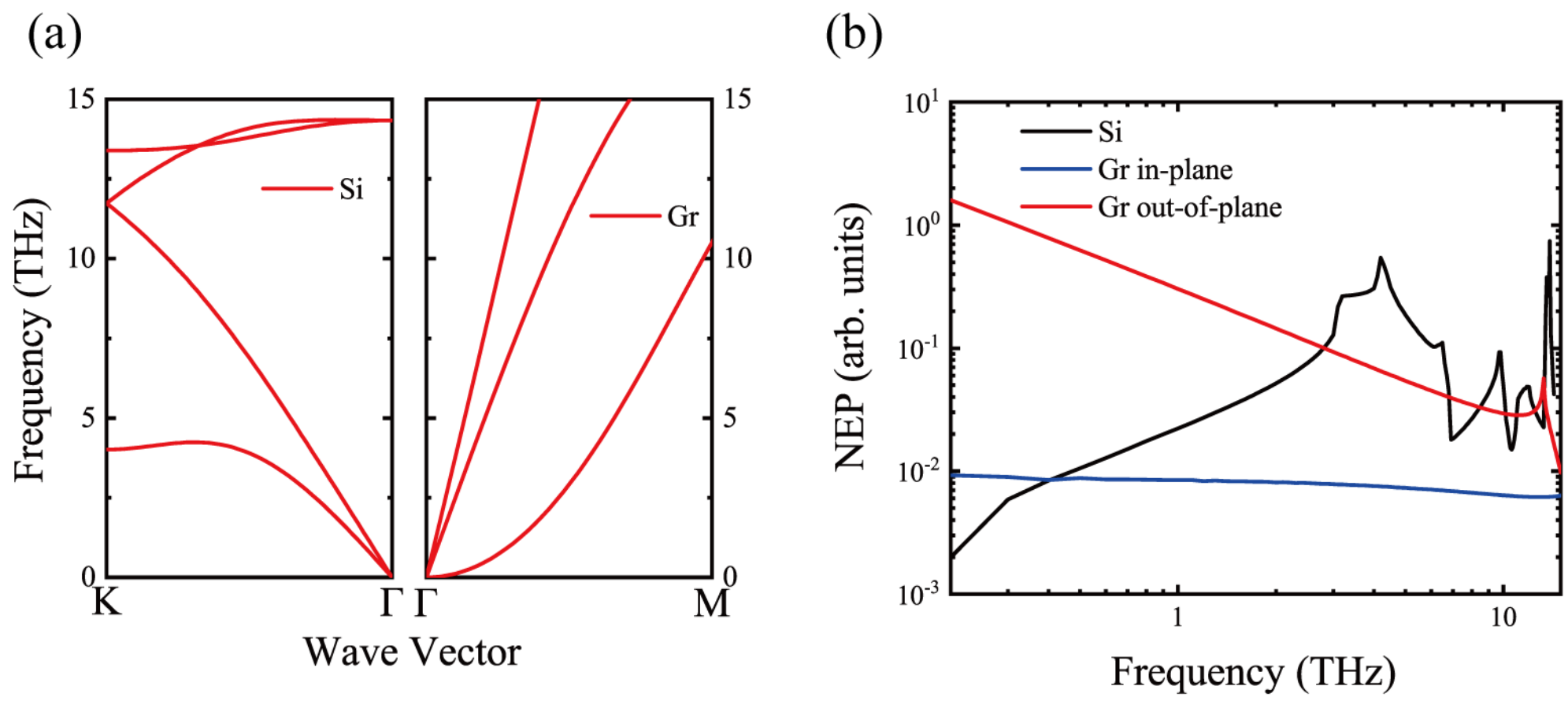

Fig. 1. (a) Phonon dispersions and (b) number of excited phonons (NEP) of typical bulk silicon (Si) and 2D graphene (Gr) materials.

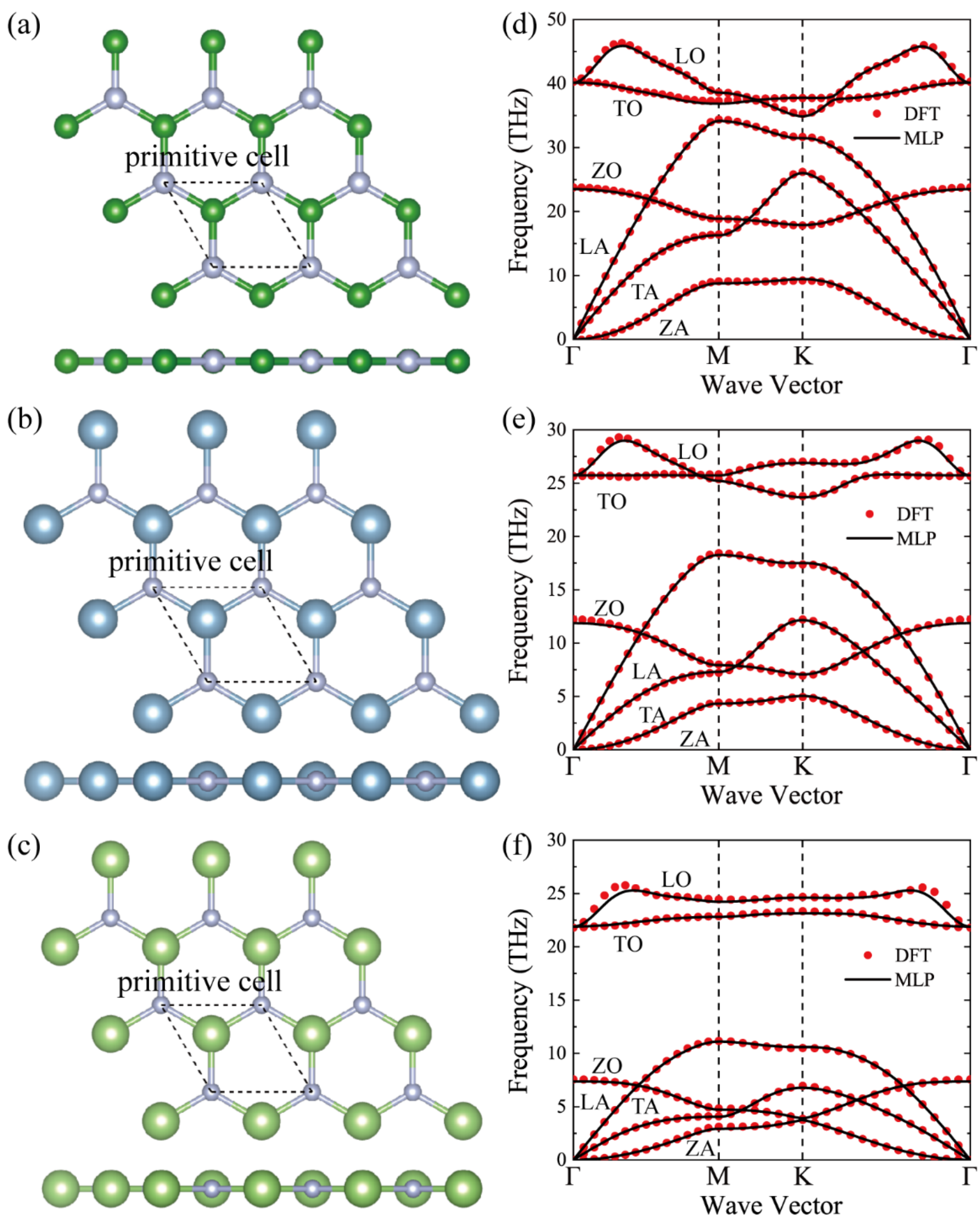

Fig. 2. (a)-(c): Top and side views of optimized monolayer (a) h-BN, (b) h-AlN, and (c) h-GaN. The green, blue, light green, and gray balls depict B, Al, Ga, and N, respectively. (d)-(f): Phonon dispersions of (d) h-BN, (e) h-AlN, and (f) h-GaN along the high-symmetry paths. The red circles and black lines are the results from first-principles and MLP, respectively.

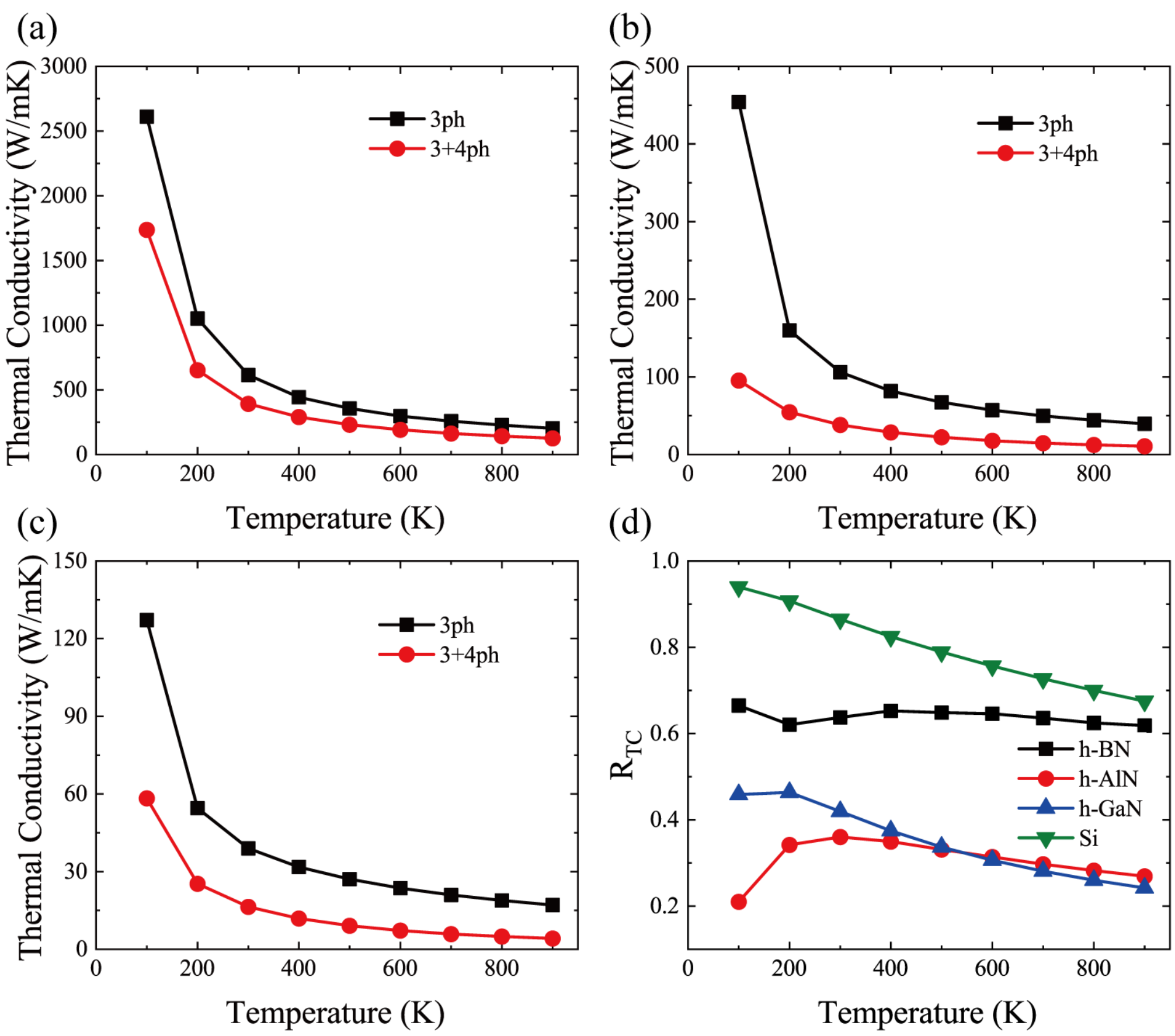

Fig. 3. (a)-(c): Thermal conductivity of (a) h-BN, (b) h-AlN, and (c) h-GaN as a function of temperature. The black line and red line indicate the results of 3ph and 3+4ph from BTE method. (d) The ratio of TC (denoted as $R_{TC}$) obtained from 3+4ph processes to that from 3ph processes for h-XN monolayers and bulk Si.

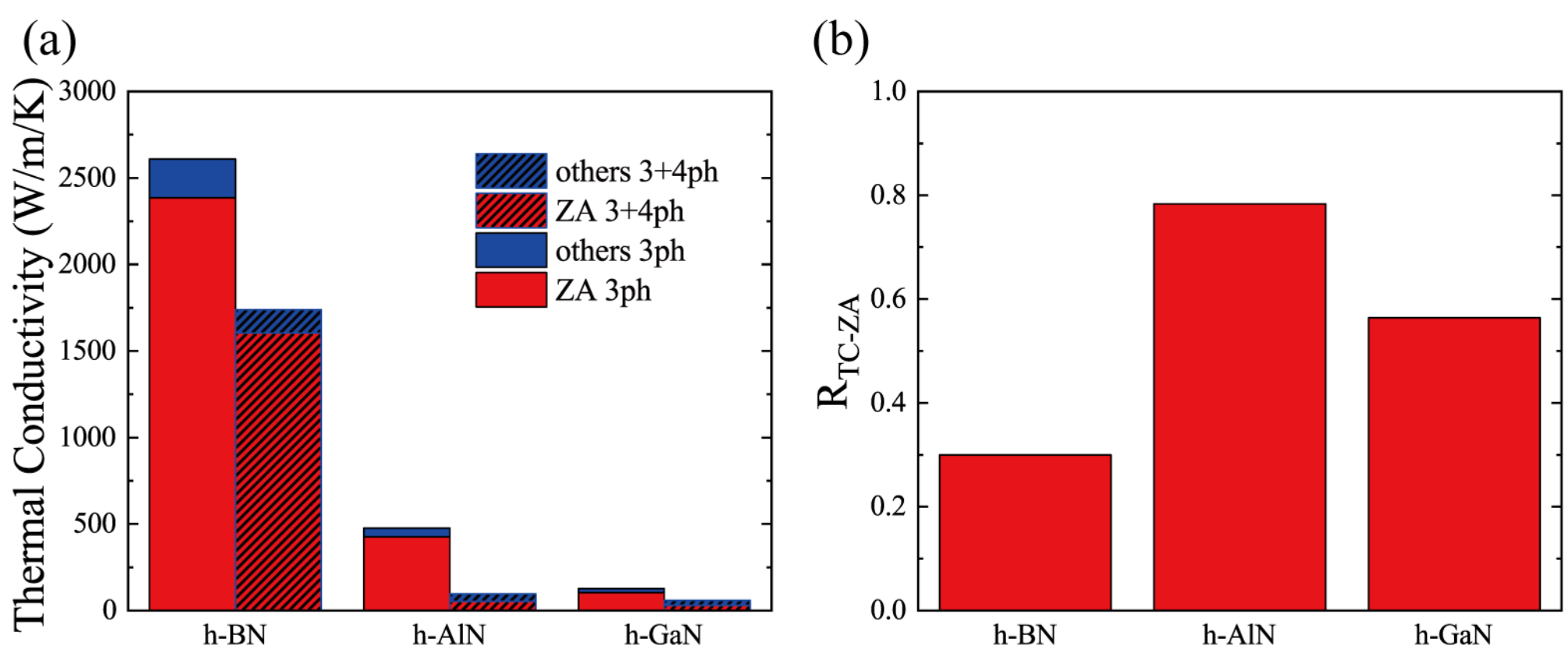

Fig. 4. (a) Thermal conductivity of multiple modes for h-XN monolayers at 100K. (b) The ratio of TC difference between 3ph and 3+4ph scatterings induced by ZA mode and $TC_{3ph}$ ($R_{TC-ZA}$), which is defined as $\Delta TC_{ZA}/TC_{3ph}$. Here $\Delta TC_{ZA} = TC_{3ph-ZA} - TC_{3+4ph-ZA}$, where $TC_{3ph-ZA}$ and $TC_{3+4ph-ZA}$ represent the TC contributed by ZA mode by considering only 3ph scattering and 3+4ph scattering, respectively.

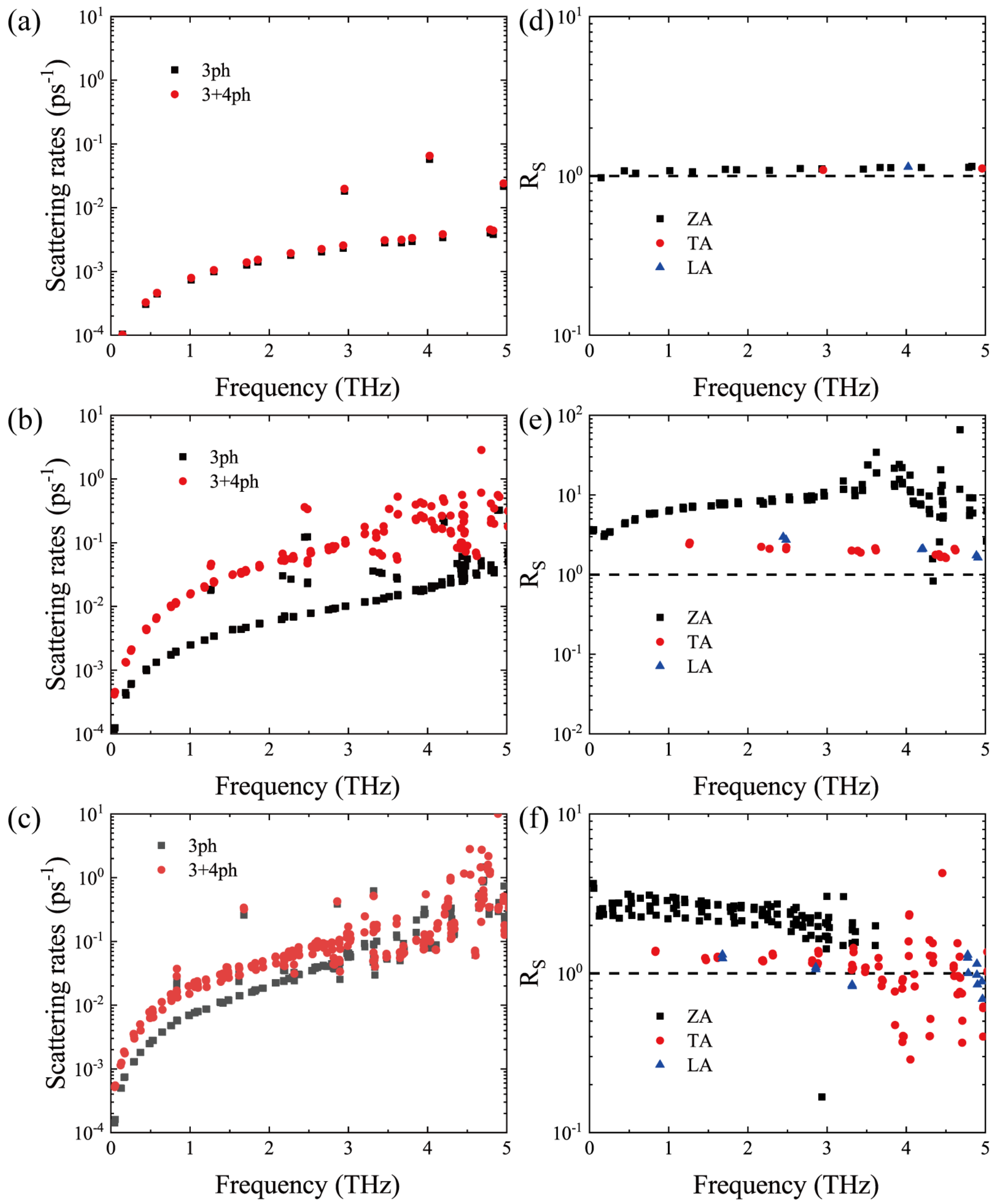

Fig. 5. (a)-(c): Scattering rates at 100 K of 3ph and 3+4ph for (a) h-BN, (b) h-AlN, and (c) h-GaN, respectively. (d)-(f): The corresponding ratio of scattering rates ($R_S$) between 3+4ph and 3ph processes for (d) h-BN, (e) h-AlN, and (f) h-GaN, respectively.